\documentclass[pra,showpacs,twocolumn,notitlepage,superscriptaddress]{revtex4-1}

\usepackage[utf8]{inputenc}
\usepackage[english]{babel}
\usepackage[T1]{fontenc}

\usepackage{graphicx}
\usepackage{amssymb,amsfonts,amsmath}
\usepackage{braket}
\usepackage{hyperref}
\usepackage{color}
\usepackage{soul}

\newcommand{\eg}[0]{\textit{e.g. }}  
\newcommand{\etal}[0]{\textit{et. al}}
\newcommand{\I}[0]{\mathcal{I}}
\newcommand{\M}[0]{\mathcal{M}}

\newcommand{\inp}[0]{\mathrm{in}}
\newcommand{\out}[0]{\mathrm{out}}
\newcommand{\qfi}[0]{F_\mathrm{Q}}
\newcommand{\ind}[0]{\mathrm{ind}}
\newcommand{\dis}[0]{\mathrm{dis}}
\newcommand{\tr}[0]{\mathrm{tr}}

\begin{document}

\title{Role of indistinguishability in interferometric phase estimation}

\author{L. T. Knoll}
\affiliation{DEILAP, CITEDEF-CONICET, J.B. de La Salle 4397, 1603 Villa Martelli,
Buenos Aires, Argentina}
\affiliation{Departamento de F\'{\i}sica, FCEyN, UBA. Ciudad Universitaria, 1428 Buenos Aires, Argentina}
\author{G. M. Bosyk}
\affiliation{Instituto de F\'isica La Plata, UNLP, CONICET, Facultad de Ciencias Exactas, La Plata, Argentina}
\affiliation{Universit\`{a} degli Studi di Cagliari, Via Is Mirrionis 1, I-09123, Cagliari, Italy}
\author{I. H. L\'opez Grande}
\affiliation{DEILAP, CITEDEF-CONICET, J.B. de La Salle 4397, 1603 Villa Martelli,
Buenos Aires, Argentina}
\affiliation{Departamento de F\'{\i}sica, FCEyN, UBA. Ciudad Universitaria, 1428 Buenos Aires, Argentina}
\author{M. A. Larotonda}
\affiliation{DEILAP, CITEDEF-CONICET, J.B. de La Salle 4397, 1603 Villa Martelli,
Buenos Aires, Argentina}
\affiliation{Departamento de F\'{\i}sica, FCEyN, UBA. Ciudad Universitaria, 1428 Buenos Aires, Argentina}

\email{Corresponding author: XXX}




\begin{abstract}

We report a theoretical and experimental study on the role of indistinguishability in the estimation of an interferometric phase.
In particular, we show that the quantum Fisher information, which limits the maximum precision achievable in the parameter estimation, increases linearly with respect to the degree of indistinguishability
between the input photons in a two-port interferometer, in the ideal case of a pure probe state. We experimentally address the role played by the indistinguishability for the case of two photons entering a polarization-based interferometer, where the degree of indistinguishability is characterized by the overlap between two spatial modes. The experimental results support the fact that, even in the presence of white noise, a quantum enhancement in the interferometric phase estimation can be obtained from a minimum degree of indistinguishability.

\end{abstract}


\date{\today}

\maketitle

\section{Introduction}

Quantum metrology allows us to obtain an increase in precision with respect to the classical framework in the problem of parameter estimation, by exploiting quantum phenomena such as entanglement or squeezing.
Beyond its fundamental role, it has been applied to many and diverse problems such as in gravitational-wave detection \cite{Aasi2013},  quantum imaging \cite{Mitchell2004,Brida2010}, atom clocks \cite{Bloom2014,Ludlow2015}, among others (see \eg the reviews \cite{Giovanetti2004,Giovanetti2011,Toth2014,Demkowicz2015,Degen2017,Pezze2018,Berchera2018} and references therein for general introduction and some of its applications).
One can resume the general procedure of quantum parameter estimation following the three basics steps: (i) preparation of
an initial quantum state, the so-called probe state, $\rho^\inp$, (ii) encoding the parameter $\phi$ to be estimated into a quantum state $\rho^\out(\phi)=\mathcal{E}_\phi(\rho^\inp)$ by means of a quantum operation $\mathcal{E}_\phi$ applied to the initial state, and (iii) performing a measurement on the final state $\rho^\out(\phi)$ that is, in general, given by a positive-operator valued measure (POVM) $\M=\{M_m\}$ with possible outcomes $m$.
Then, an estimator $\hat{\phi}$ can be obtained from the probabilities of the outcomes that are, according to the Born rule, $p(m|\phi)=\tr[M_m \rho^\out(\phi)]$.
Clearly, the variance of the estimator $\Delta^2(\hat{\phi})$ depends on the chosen probe state, the quantum operation that encodes the parameter and the performed measurement.
Although exploiting quantum phenomena can enhance the parameter estimation, there is an ultimate limit in the precision given by the so-called quantum Cram\'er-Rao bound~\cite{Helstrom1967,Holevo1982} (QCRB)
\begin{equation}
\label{eq:QCRB}
\Delta^2(\hat{\phi}) \geq \frac{1}{F_Q(\phi)},
\end{equation}
where $F_Q(\phi) = \max_{\{\M\}} F(\phi)$ is the quantum Fisher information (QFI), with
\begin{equation}
\label{eq:FI}
F(\phi)=\sum_m \frac{1}{p(m|\phi)}\left|\frac{dp(m|\phi)}{d\phi}\right|^2,
\end{equation}
the Fisher information (FI) of the corresponding probabilities $p(m|\phi)=\tr[M_m \rho^\out(\phi)]$ associated to a POVM $\M$ for the given state $\rho^\out(\phi)$, and the maximum is taken over all possible POVMs ${\{\M\}}$.
In this sense, QFI gives the maximal information on the parameter encoded in the state $\rho^\out(\phi)$.
Conditions for the saturation of the QCRB~\eqref{eq:QCRB} have been studied in the seminal works~\cite{Helstrom1967,Holevo1982,Braunstein1994}, but these conditions depends, in general, on a measurement scheme that requires knowledge about the value of the parameter to be estimated.

For the task of estimating a phase $\phi$ from the unitary map $U_\phi = \exp(-i\phi H)$, where $H$ is the corresponding generating hermitian operator (the ``hamiltonian''), the QFI does not depend on $\phi$.
Moreover, for initial pure states $\rho^\inp = \ket{\psi^\inp}\bra{\psi^\inp}$, the QFI is proportional to the variance of the hamiltonian in the initial state, that is, $F_Q(\phi) \equiv \qfi(\ket{\Psi^\inp},H) = 4 \Delta^2(H)$ with $\Delta^2(H) =\braket{\Psi^\inp |H^2|\Psi^\inp} - \left(\braket{\Psi^\inp |H|\Psi^\inp}\right)^2$ (see \eg \cite{Paris2009,Toth2014}).
For this configuration, it has recently been shown that it is possible to saturate the QCRB with a projective measurement that does not require any previous information about the value of the parameter to be estimated~\cite{Toscano2017}.
In this regard, QFI plays a fundamental role as a witness of the maximum precision achievable.

In this paper, we focus on the problem of estimating an unknown phase $\phi$ by analyzing a general scenario of encoding it in a quantum system of a definite number of photons ($2n$) by means of a unitary map $U_\phi = \exp(-i\phi J_y)$, with $J_y = -i(\alpha^\dag \beta - \alpha \beta^\dag)/2$, where the annihilation and creation operators $\alpha,\beta,\alpha^\dag$ and $\beta^\dag$ satisfy the usual bosonic commutation relations $[\alpha,\alpha^\dag] = [\beta,\beta^\dag]=  1$.
In general, this unitary map describes any two-port interferometer with input modes $\alpha$ and $\beta$, like the ubiquitous Mach-Zehnder interferometer (see \eg \cite{Yurke1986,Kim1998,Sanders1995,Pezze2006}).
If the $2n$ photons enter the interferometer through one of the ports, whereas the other one is fed by the vacuum state, the probe state is given by $\ket{\psi^\inp} = \ket{2n,0}$ and the QFI is equal to  $F_Q(\ket{\psi^\inp},J_y)=2n$.
This leads to the standard quantum limit (SQL) scaling $\Delta^2(\hat{\phi})= 1/2n$.
It is well-known that a better choice of the probe state can surpass the SQL.
Indeed, it has been shown in~\cite{Lang2014} that the maximal QFI over initial product states with definite total photon number ($2n$) is given by $F_Q(\ket{\psi^\inp},J_y)=2n(n+1)$, and the probe state that attains this maximal value is the so-called Holland-Burnett state~\cite{Holland1993} (or twin-Fock state): $\ket{\psi^\inp} = \ket{n,n}$.
In this situation, one can obtain a phase error $\Delta^2(\hat{\phi})\sim 1/2n^2$ that scales as the Heisenberg-limit (HL) (see \eg \cite{Bollinger1996,Lee2002}).
We notice that attaining this limit is possible due to the quantum interference between $2n$ completely indistinguishable photons entering into the interferometer.
In a more realistic scenario, this precision might be limited by the presence of an extra unavoidable or uncontrollable degree of freedom that degrades the indistinguishability of interfering photons.
Since achieving complete indistinguishability between interfering photons can be a technological challenge, the characterization of such effect naturally arises.
This point has been theoretically and experimentally addressed in \cite{Birchall2016} for small changes of the phase, where the precision of the estimation characterized by FI is studied as a function of a degree of indistinguishability due to a path delay mismatch between interfering photons.
Indeed, for local phase estimation, it has been shown that this undesirable effect can be mitigated using two photons with a carefully-engineered accessible additional degree of freedom \cite{Jachura2016}.
Here, we do not restrict to small changes of phase and we do not assume any previous knowledge about the phase value.
We are interested in characterizing how the precision of the estimation, quantified by the QFI, is affected by a degree of indistinguishability between interfering photons in a general scenario.
To measure this effect, we designed an experimental setup for the case of two photons entering a polarization-based interferometer, where the degree of indistinguishability is characterized and controlled by the overlap between two spatial modes.
Our results support the fact that indistinguishability is essential to obtain a quantum enhancement in the interferometric phase estimation.
Such description complements other studies, in addition to the later ones \cite{Birchall2016,Jachura2016}, of non-ideal scenarios, for example those addressed in Refs~\cite{Demkowicz2009,Datta2011,Escher2011,Demkowicz2012,Roccia2018}.

\section{Quantum Fisher information vs indistinguishability}
\subsection{General scheme}

\begin{figure}[h!]
\vspace{0.5cm}
  \centering
  \includegraphics[width=0.47\textwidth]{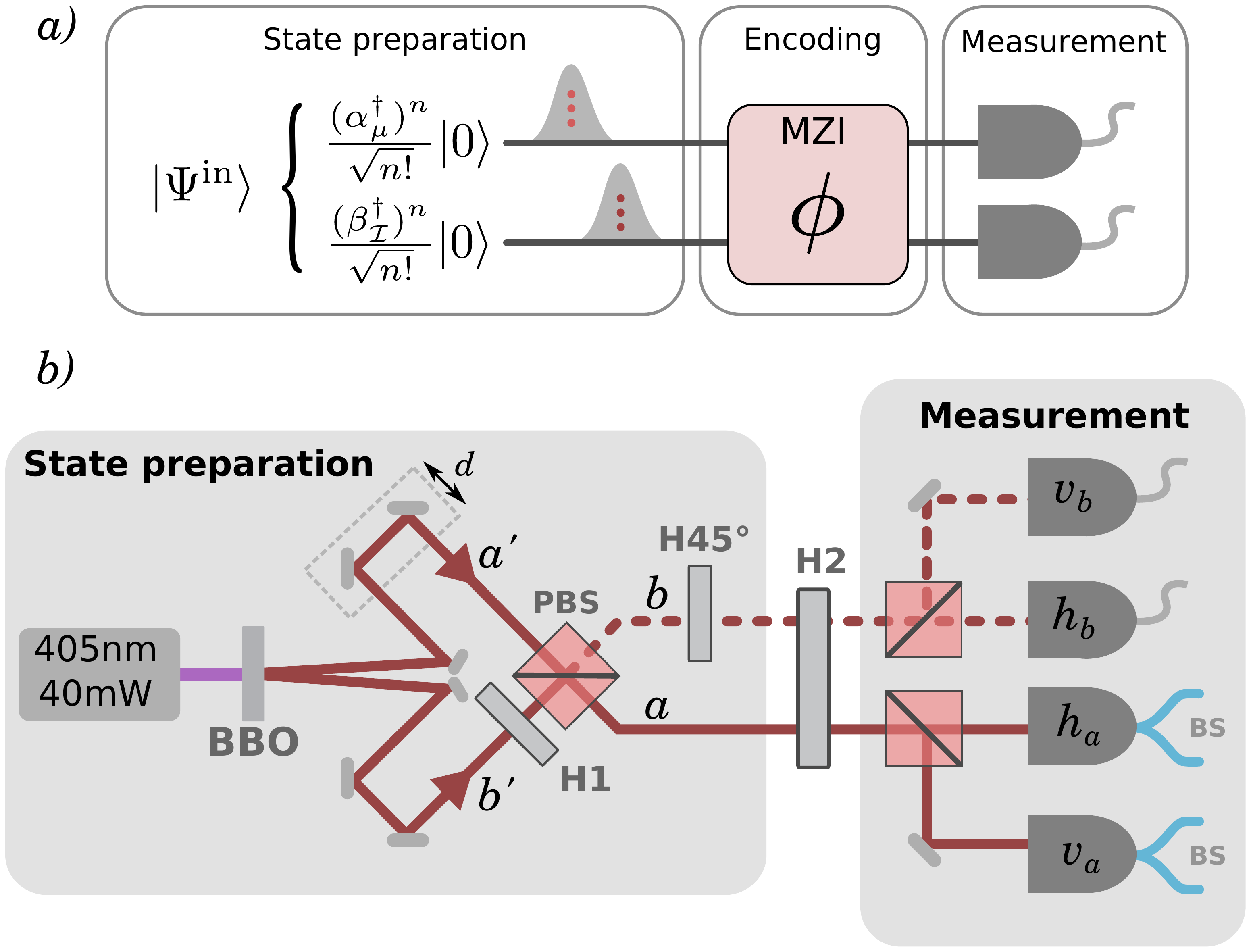}
  \caption{(a) General scheme for studying the role of indistinguishability ($\I$) of photons in a two-port interferometer to estimate the phase ($\phi$).
  (b) Experimental setup. Horizontally polarized photon pairs are generated by SPDC into two path modes $a'$ and $b'$. State preparation is perferomed by using half-wave plate (HWP) H1 in path $b'$, the action of the polarizing beam splitter (PBS) and an additional HWP at $45^{\circ}$ at output path mode $b$. HWP H2 acts as a polarization interferometer, adding a relative phase between polarization modes. Measurements are performed in coincidence between the four possible output modes $h_a,v_a,h_b,v_b$. Each output mode is coupled into a single mode fiber. Alternatively, fiber beam splitters are added to measure double photon events (see text for details).}
  \label{fig:setup}
\end{figure}

Let us consider the general scheme of Fig.\ref{fig:setup}a) to estimate an unknown phase $\phi$ encoded in a state of $2n$ photons through the action of a two-port interferometer.
To take into account a degree of indistinguishability between interfering photons of different ports, we consider a four mode representation of individual photons.
More precisely, let $\alpha_\mu, \alpha_\nu, \beta_\mu$ and $\beta_\nu$ be annihilation operators, where the labels $\alpha$ and $\beta$ refer to input ports of the interferometer, while $\mu$ and $\nu$ correspond to an extra degree of freedom that allows photons to be distinguished.
The annihilation and creation operators satisfy usual bosonic commutation relations $[\alpha_\mu,\alpha^\dag_\nu] = [\beta_\mu,\beta^\dag_\nu]=  \delta_{\mu\nu}$.
The degree of indistinguishability, $0\leq \I \leq 1$, between interfering photons of different ports is then modelled by assuming that $n$ photons occupy the mode $\alpha_\mu$, whereas the other $n$ are in a superposition of orthogonal modes $\beta_\mu$ and $\beta_\nu$, that is, $\beta_\I = \sqrt{\I} \beta_\mu + \sqrt{1-\I} \beta_\nu$ (in a similar fashion to Refs.\cite{Ra2013a,Ra2013b} where $\mu$ and $\nu$ correspond to temporal modes).
In this way, the initial state can be written as
\begin{widetext}
\begin{equation}
 \begin{aligned}
\label{eq:input}
  \ket{\Psi^\inp} & = \frac{(\alpha_\mu^\dag)^n}{\sqrt{n!}} \frac{(\beta_\I^\dag)^n}{\sqrt{n!}} \ket{0}\\
   & = \I^{\frac{n}{2}} \ket{n_{\alpha}n_{\beta}00} + \sum_{k=1}^{n-1} \sqrt{\frac{n!}{k!(n-k)!}} \I^{\frac{n-k}{2}} (1-\I)^{\frac{k}{2}} \ket{n_{\alpha}(n-k)_{\beta}0k_{\beta}} + (1-\I)^{\frac{n}{2}} \ket{n_{\alpha}00n_{\beta}}.
\end{aligned}
\end{equation}
\end{widetext}
We are using the notation $\ket{n_{\alpha}n_{\beta}00} \equiv \ket{n_{\alpha}n_{\beta}}_{\mu}\otimes\ket{00}_{\nu}$, where the first ket refers to $2n$ photons in mode $\mu$, whereas the second one indicates zero photons in mode $\nu$.
Notice that the initial state is, in general, entangled with respect to the mode bipartition $(\mathcal{\mu},\mathcal{\nu})$ formed from the corresponding subalgebras of the modes $\mu$ and $\nu$, but not with respect to the mode bipartition $(\alpha,\beta)$ formed from the corresponding subalgebras of modes $\alpha$ and $\beta$ (see \eg\cite{Benatti2010} for a discussion about entanglement of identical particles).
In particular, when $\I =1$ the $2n$ input photons are completely indistinguishable and the initial state takes the form $\ket{\Psi^\ind} = \ket{n_{\alpha}n_{\beta}00}$.
Conversely, when $\I=0$, the input photons on different ports are completely distinguishable and the initial state is given by $\ket{\Psi^\dis} = \ket{n_{\alpha}00n_{\beta}}\equiv \ket{n_{\alpha}0}_{\mu}\otimes\ket{0n_{\beta}}_{\nu}$.

The hamiltonian responsible for encoding the phase $\phi$ into the initial state is given by $H= J_\mu \otimes I + I \otimes J_\nu$, where $J_\chi = -i(\alpha_\chi^\dag \beta_\chi - \alpha_\chi \beta^\dag_\chi)/2$ with $\chi=\mu,\nu$.
Accordingly, the quantum Fisher information for this scheme is given by
\begin{equation}
\label{eq:QFI}
  \qfi(\ket{\Psi^\inp}, H) =  2n^2 \I + 2n.
\end{equation}
The proof of this result is given in App.\ref{app:QFIproof}.
Therefore, we find that the QFI increases linearly  with the degree of indistinguishability.
Moreover, the QFI \eqref{eq:QFI} can be written as a convex combination of two extreme cases, that is, $\qfi(\ket{\Psi^\inp}, H)= \I \qfi(\ket{\Psi^\ind}, H) + (1-\I) \qfi(\ket{\Psi^\dis}, H)$ where $\qfi(\ket{\Psi^\ind}, H)=2n(n+1)$ and $\qfi(\ket{\Psi^\dis}, H)=2n$.
We also notice that $\qfi(\ket{\Psi^\ind}, H)=2n(n+1)$ is equal to the QFI of the Hollan-Burnett states that scales as the HL.
On the other hand, $\qfi(\ket{\Psi^\dis}, H)=2n$ is equal to the QFI of the $2n$ photons entering the interferometer on one of the ports (while the other port is fed by the vacuum state) and leads to the SQL scaling.
Interestingly, these results imply that, in the ideal case, one can beat the SQL with any no-null degree of indistinguishability between the $2n$ interfering photons.
In other words, indistinguishability is a necessary and sufficient condition in order to obtain a quantum advantage in the estimation of an interferometric phase in this scheme.

Additionally, in a more realistic scenario, the quantum state is not generally a perfect and pure state.
It is reasonable to assume a model where the state preparation is affected by white noise, independently of the degree of indistinguishability $\I$.
In this situation, the initial state can be modeled as a mixture between $\ket{\Psi^\inp}$ given by Eq.~\eqref{eq:input} and the maximally mixed state $\mathbb{I}/d$; that is,
\begin{equation}
\label{eq:inputmixed}
\rho^\inp= (1-\epsilon) \ket{\Psi^\inp}\bra{\Psi^\inp} + \epsilon \frac{\mathbb{I}}{d},
\end{equation}
where $\epsilon$ is the degree of white noise and $d=(2n+3)(2n+2)(2n+1)/3!$ is the dimension of the Hilbert space of a four-mode system of $2n$ identical photons.
In this way, the QFI given in \eqref{eq:QFI} is reduced by a factor that depends on the degree of noise as follows,
\begin{equation}
 \begin{aligned}
\label{eq:QFImixed}
  \qfi(\rho^\inp, H) =& \frac{(1-\epsilon)^2}{1-(1-\frac{2}{d})\epsilon} \qfi(\ket{\Psi^\inp}, H)\\
  =& \frac{(1-\epsilon)^2}{1- (1-\frac{2}{d})\epsilon}  (2n^2 \I + 2n).
\end{aligned}
\end{equation}
Unlike the pure state case, a minimum degree of indistinguishability given by $\I_{\min} = \frac{(2+d(1-\epsilon))\epsilon}{nd(1-\epsilon)^2}$ is now required to surpass the SQL, as long as the degree of noise remains below $\epsilon_{\max} = \frac{2 + (2n+1)d - \sqrt{4 + 4 d + d^2 + 8 nd}}{2 (n+1)d}$. Therefore, indistinguishability is still a necessary condition but not sufficient in order to surpass de SQL.

\subsection{Two-photon state in a two-port interferometer}

To experimentally study the role of indistinguishability, let us consider the setup of Fig.\ref{fig:setup}b), consisting of a two-photon state ($2n=2$) entering a two-port polarization-based interferometer.
In this setup, the polarization modes $h$ (horizontal) and $v$ (vertical) play the role of modes $\alpha$ and $\beta$ of the general scheme discussed above, whereas the spatial modes $a$ (lower output port) and $b$ (upper output port) play the role of $\mu$ and $\nu$.

The initial two-photon state is prepared as follows.
Pairs of correlated photons with horizontal polarization are generated by spontaneous parametric down-conversion (SPDC) in a BBO non-linear crystal into two path modes which we call $a'$ and $b'$.
Photons in path $b'$ pass through a controlled half-wave plate (HWP) that rotates their polarization into a superposition of horizonal and vertical modes, that is, $h \mapsto \cos(2\varphi) h+\sin(2\varphi) v$ (H1 in Fig.\ref{fig:setup}b)) where $\varphi$ is the physical angle of the HWP.
Subsequently, both photons enter a polarizing beam splitter (PBS). The optical path $a'$ is adjusted by using a translation stage so that both photons arrive at the PBS at the same time.
The action of the polarizing beam splitter is such that horizontally polarized input photons from path $a'$ are transmitted unchanged into path mode $a$. On the other hand, input photons from path $b'$ with horizontal polarization are transmitted into path mode $b$, whereas photons with vertical polarization are reflected into path mode $a$.
Placing a half-wave plate at $45^\circ$ (H$45^\circ$ in Fig.\ref{fig:setup}b)) in the output path mode $b$ changes the horizontally polarized photons into vertical ones.
Therefore, the initial horizontally polarized photons in path $b'$ are prepared in a superposition of spatial modes $a$ and $b$ with vertical polarization, that is, $h_{b'}\mapsto v_{b'} = \sin(2\varphi) v_a + \cos(2\varphi) v_b$. This allows us to prepare the following initial two-photon state

\begin{equation}
\begin{aligned}
\label{eq:initial2phtons}
 \ket{\Psi^\inp} = &(h_{a'})^\dag (v_{b'}^\dag) \ket{0} \\
 =& \sqrt{\I(\varphi)} \ket{1_h1_v00} + \sqrt{1-\I(\varphi)} \ket{1_h001_v},
\end{aligned}
\end{equation}
where the degree of indistinguishability of interfering photons is controlled by the angle $\varphi$ of half-wave plate H1 and it is given by $\I(\varphi) = \sin^2 (2\varphi)$.

A phase $\phi$ is encoded into the initial state by the action of another controlled half-wave plate (H2 in Fig.\ref{fig:setup}b)), which acts as a polarization interferometer. This HWP performs the transformation of the polarization modes $h_a \mapsto \cos(2\theta) h_a+\sin(2\theta) v_a$ and $v_i \mapsto \sin(2\theta) h_i - \cos(2\theta) v_i$ with $i=a,b$ and where $\theta=\phi/4$ is the physical angle of the HWP.
Therefore, the state at the output of the interferometer takes the form
\begin{widetext}
\begin{equation}
\begin{aligned}
\label{eq:out2phtons}
 \ket{\Psi^\out} & = \sqrt{\I(\varphi)}  \left[\frac{\sqrt{2}\sin(\phi)}{2}(\ket{2_h000} - \ket{02_v00}) - \cos(\phi) \ket{1_h1_v00}
 \right] \\
   & + \sqrt{1- \I(\varphi)} \left[\frac{\sin(\phi)}{2} \left( \ket{1_h01_h0} - \ket{01_v01_v} \right)- \cos^2(\phi/2)\ket{1_h001_v} +\sin^2(\phi/2)\ket{01_v1_h0}\right].
\end{aligned}
\end{equation}
\end{widetext}

The QFI is obtained by measuring the optimal probabilities $\{p(m|\phi) = |\braket{m|\Psi^\out}|^2\}_{m \in \mathfrak{M}}$  through the projective measurement $\mathcal{M}=\{\ket{m}\bra{m}\}$, where $\mathfrak{M} = \{2_h000,02_v00,1_h1_v00,1_h01_h0,01_v01_v,1_h001_v,01_v01_v, \allowbreak 001_h1_v,002_h0,0002_v\}$ (although projections onto the last three elements do not contribute to the FI, see App.~\ref{app:prob}), which gives
\begin{equation}
\label{eq:QFI2}
  \qfi(\ket{\Psi^\inp}, H) =  \sum_{m \in \mathfrak{M}} \frac{\left|\frac{dp(m|\phi)}{d\phi}\right|^2}{p(m|\phi)}= 2 \I(\alpha) + 2.
\end{equation}

Projections onto polarization modes $h,v$ are implemented with a PBS at each output path $a,b$ and single-photon counting modules are placed at each output mode corresponding to $h_a,v_a,h_b,v_b$. Coincidence measurements are performed between the corresponding detectors according to each probability measurement. Each output mode is coupled into a single mode fiber and 10nm interference filters are placed before the optics corresponding to exit modes $h_a,v_a$ while band-pass filters are used for outputs $h_b,v_b$.
Projective measurements onto $\{2_h000,02_v00\}$ are obtained including fiber beamsplitters (BS) at the output modes $h_a$ and $v_a$ as shown Fig.\ref{fig:setup}b) and measuring coincidences between the outputs of each BS.

Let us consider the case where the state preparation is affected by white noise.
Accordingly, the initial state is of the form~\eqref{eq:inputmixed}, that is,  $\rho^\inp= (1-\epsilon) \ket{\Psi^\inp}\bra{\Psi^\inp} + \epsilon \frac{I}{10}$ with  $\ket{\Psi^\inp}$ given by Eq.~\eqref{eq:initial2phtons}.
For the same projective measurement $\M$, the probabilities change as $p'(m|\phi)=(1-\epsilon)p(m|\phi)+\epsilon\frac{1}{10}$.
It can be shown that the Fisher information from these noisy probabilities gives
\begin{widetext}
\begin{align}
\label{eq:FIruido}
  F'(\phi) &=
  \frac{4(1-\I)^2(1-\epsilon)^2 \cos^6(\phi/2) \sin^2(\phi/2)}{\epsilon/10 + (1 - \I) (1 - \epsilon) \cos^4(\phi/2)}
  +  \frac{4(1-\I)^2(1-\epsilon)^2 \cos^2(\phi/2) \sin^6(\phi/2)}{\epsilon/10 + (1 - \I) (1 - \epsilon) \sin^4(\phi/2)} \\ \nonumber
  &+  \frac{4\I^2(1-\epsilon)^2 \cos^2(\phi) \sin^2(\phi)}{\epsilon/10 + \I (1 - \epsilon) \cos^2(\phi)} +
  \frac{(1-\I)^2(1-\epsilon)^2 \cos^2(\phi) \sin^2(\phi)}{\epsilon/5 + (1 - \I) (1 - \epsilon) \sin^2(\phi)/2} 
  + \frac{2\I^2(1-\epsilon)^2 \cos^2(\phi) \sin^2(\phi)}{\epsilon/10 + \I (1 - \epsilon) \sin^2(\phi)/2} . \nonumber
\end{align}
\end{widetext}
and it is strictly lower than the corresponding QFI,
$\qfi(\rho^\inp, H) = \frac{(1-\epsilon)^2}{1-\frac{4}{5}\epsilon}  (2\I + 2),$ (c.f. Eq.~\eqref{eq:QFImixed} with $n=1$ and $d=10$)
for any amount of noise.
This means that considering an initial mixed state, the measurement $\M$ is not the optimal one for any value of $\I$ and $\phi$.
However, we will see that for a reasonable small amount of noise given by $\epsilon=0.06$, the difference between $\max_\phi F'(\phi)$ and $\qfi(\rho^\inp, H)$ is of the order of $7\%$ in the worst case. Therefore, $\max_\phi F'(\phi)$ is a good approximation of the QFI for this case.

\subsection{Experimental results}

\begin{figure}[htbp!]
  \centering
  \includegraphics[width=0.45\textwidth]{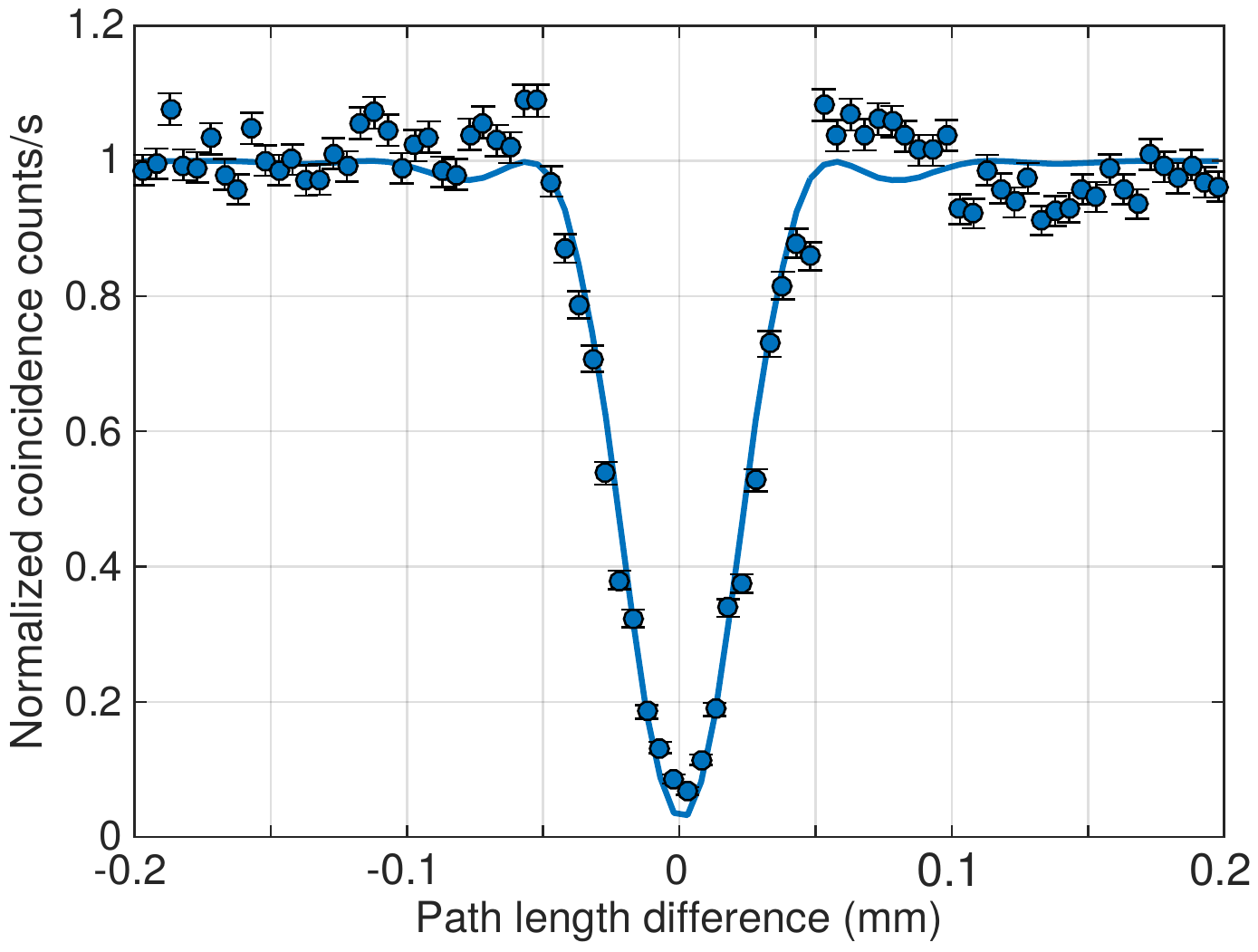}
  \caption{Normalized coincidence counts as a function of the path length difference between interfering photons. The solid line corresponds to the Fourier transform of the 10nm interference filters.}
  \label{fig:hom}
\end{figure}

As mentioned in the previous section, the optical path length of mode $a'$ is adjusted so that both photons overlap at the PBS. Coincidence counts are measured at the output with the setting $\varphi=45^{\circ}$ and $\phi=\frac{\pi}{2}$ ($\theta=\frac{\pi}{8}$), so as to observe the interference effect.
This is essentially equivalent to the traditional Hong-Ou-Mandel interferometer \cite{hong1987} but using polarization instead of spatial modes \cite{Birchall2016,Slussarenko2017,Matthews2016}. Therefore, by measuring the coincidences between detectors $h_a$ and $v_a$ as a function of the path length difference we observe the two-photon interference effect and identify the path length which guarantees maximum indistinguishability. We achieved an interference visibility of $\mathcal{V}=(87\pm 1)\%$. Figure \ref{fig:hom} shows the normalized coincidence counts as a function of the path length difference. The solid line corresponds to the Fourier transform of the 10nm interference filters, whose bandwidth matches the coherence length measured from the two photon interference pattern.

The degree of indistinguishability is set by the angle $\varphi$ of H1 and then corrected by the maximum indistinguishability obtained in the two photon HOM interferometer, defined from the minimum of the normalized HOM-dip as $\I_{\max}=1-\text{min(HOM)}=0.93$.
Therefore, $\I(\varphi) = \I_{\max}\sin^2 (2\varphi)$.

\begin{figure*}[htbp!]
\begin{center}
  \includegraphics[width=\textwidth]{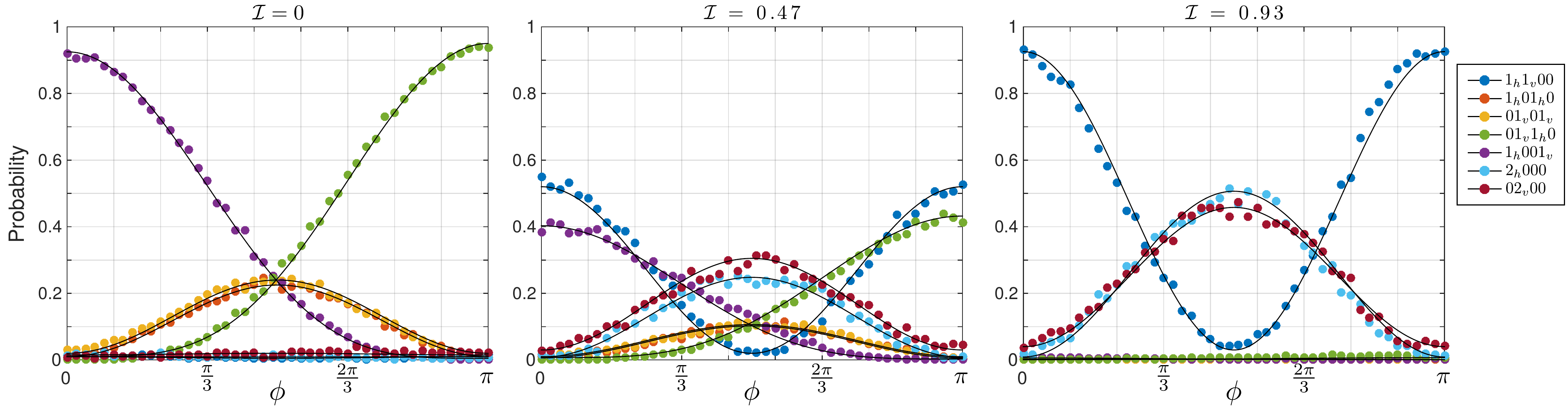}
  \caption{Experimental data: normalized coincidence count rates for different projections. The probabilities $p(m|\phi)$ needed to calculate the FI are obtained fitting these measurements with each theoretical probability (App.~\ref{app:prob}) and obtaining a proportional parameter and a constant parameter from the fit.}
  \label{fig:probas}
\end{center}
\end{figure*}

\begin{figure}[h!]
  \centering
  \includegraphics[width=0.45\textwidth]{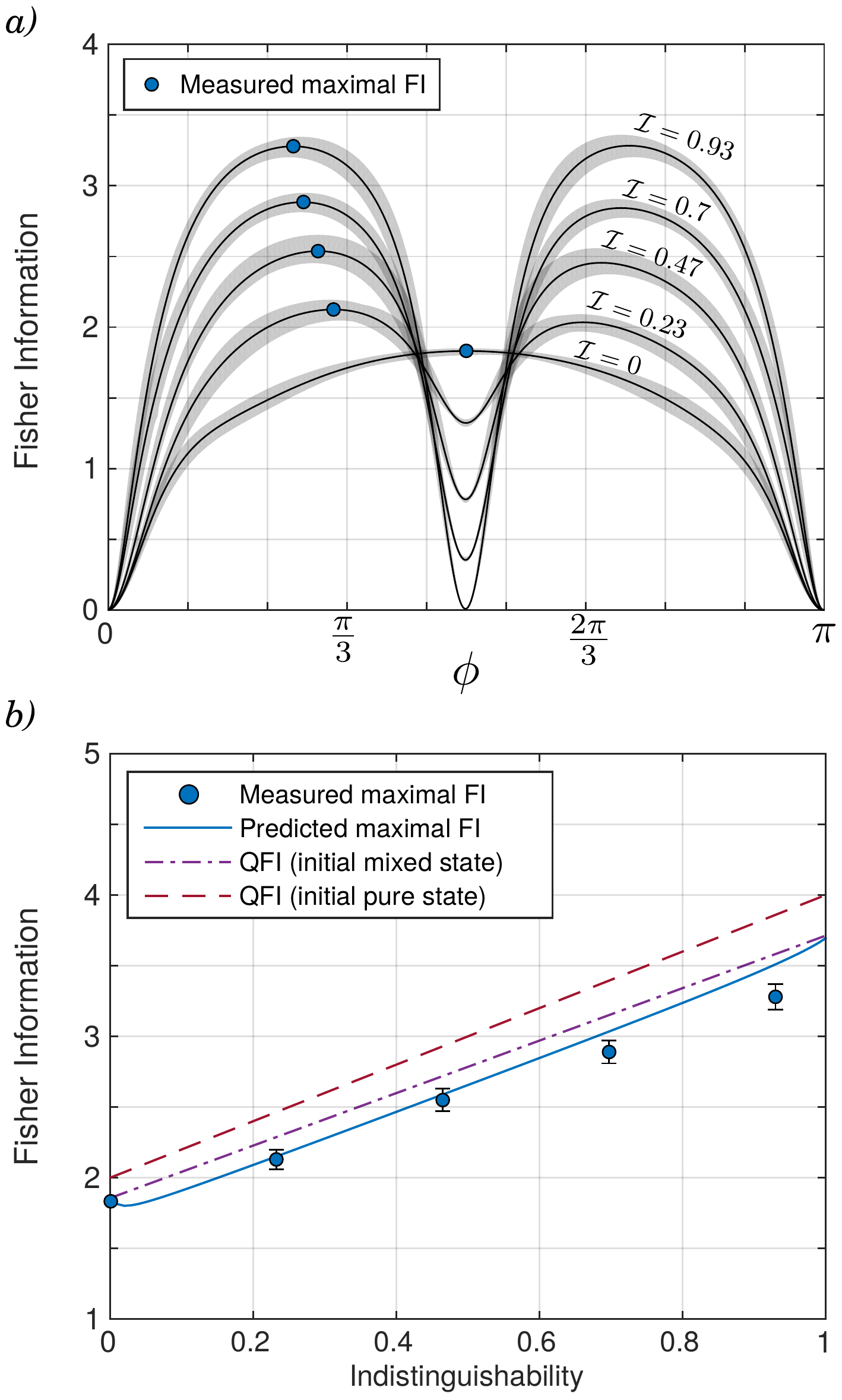}
  \caption{a) Fisher information from the fitted probabilities as a function of the phase $\phi$ for five different values of indistinguishability, in increasing order. The shaded area corresponds to the error estimated by Monte-Carlo simulations. The symbols correspond to the largest FI value for each degree of indistinguishability. b) Symbols: largest FI values as shown in a).
  Solid line: theoretical prediction assuming non-ideal probability distributions $p'(m|\phi)$ with a degree of noise $\epsilon=0.06$.
  Dash-dot line: theoretical prediction of the QFI for a mixed initial state with the same $\epsilon=0.06$. Dashed line: QFI for an initial pure state.}
  \label{fig:fisher}
\end{figure}

The different projections onto $\M$
were measured as a function of the phase $\phi$ for five different values of $\I(\varphi)=\{0, 0.23,    0.47,    0.70,   0.93 \}$. Figure \ref{fig:probas} shows normalized coincidence rates for $\I=\{0,0.47,0.93 \}$. We obtain probability distributions by fitting each measurement with a model $\text{A}p(m|\phi)+\text{B}$, where $p(m|\phi)$ are the theoretical probabilities that correspond to each projection (see App.~\ref{app:prob}). Measurements onto $\{001_h1_v,002_h0,0002_v\}$ take constant values (background photons) for all $\I$ and $\phi$ and are not shown as they do not contribute to the Fisher information. The FI is then calculated from the fitted probabilities using \eqref{eq:FI} (Figure \ref{fig:fisher}a)).
Given that these experimentally obtained probabilities are not ideal (in particular, there is a non-zero background leading to visibilities smaller than unity), we observe a dependence of the Fisher information on the phase $\phi$ as shown in Figure \ref{fig:fisher}a). This dependence on the phase has already been reported elsewhere (see for example \cite{Matthews2016}). It is worth to note that the angular dependence should disappear when the probe state is given by the pure state \eqref{eq:initial2phtons}, in the ideal noiseless  case.
There are several factors contributing to this behavior, given that the state preparation, phase encoding and measurement are all susceptible to experimental imperfections. We will restrict our analysis only to the error introduced due to imperfect preparation of the initial two-photon state generated by SPDC.
We estimate the value of $\epsilon$ (degree of impurity) as follows: we assume that for $\I=0$, which is the experimental state that can be more accurately obtained, the maximum value of the experimentally obtained FI is equal to the value given by the expression \eqref{eq:FIruido} for the equivalent conditions. From this condition we obtain $\epsilon=0.06$.
Figure \ref{fig:fisher}b) shows the theoretical prediction (solid line) calculated from the maximum of \eqref{eq:FIruido} as a function of the indistinguishability, together with the experimentally obtained FI as the maximum values of each curve in Fig. \ref{fig:fisher}a). Error bars were calculated by Monte-Carlo simulation of experimental runs with the same Poissonian statistics. The dashed line corresponds to the QFI of the initial pure state case \eqref{eq:QFI2} while the dash-dot line corresponds to the QFI of a mixed input state for $n=1$ and $\epsilon=0.06$ \eqref{eq:QFImixed}.
As mentioned before, the largest value for the Fisher information computed from the probabilities $p'(m|\phi)$ does not agree with the QFI in \eqref{eq:QFImixed} for the same value of $\epsilon$, as can be observed by the difference between the QFI curve for an initially mixed state and the predicted maximum FI curve. Interestingly, for both extreme cases $\I = 0$ and $\I = 1$ both values approach each other with a difference smaller than $1\%$, whereas for $\I = 0.1$ the largest difference of $7\%$ is attained.
We attribute the difference between the calculated and measured data for large values of the indistinguishability to the fact that our model only considers the error introduced by an imperfect input state, while other experimental imperfections become more relevant with increasing indistinguishability.

\section{Concluding remarks}

In this work, we have developed a theoretical and experimental study to characterize how the precision of a parameter estimation, quantified by the QFI, is affected by the degree of indistinguishability $\I$ between interfering photons.
The cases of a pure state and a mixed probe state have been considered.
We have obtained a linear increase of the QFI with the degree of indistinguishability and also we have found the optimal  measurement required to achieve the QFI for the pure state case.
This linear dependence is also obtained for the mixed probe case, where the QFI is reduced by a factor that depends on the amount of noise.
A discussion on the optimal measurement for the mixed probe state case can be found on Appendix \ref{app:prob}.

In our experiment, the degree of indistinguishability between two photons entering a polarization interferometer is characterized by the overlap between spatial modes.
Projective measurements were performed onto the optimal measurement base for pure input states, obtaining the probability distributions as a function of the encoded phase. The Fisher information was then calculated from these probability distributions and the experimental data was modeled taking into account a non ideal (mixed) probe state, obtaining a good agreement for $\I<0.7$.
Our results support the fact that indistinguishability is essential to obtain a quantum enhancement in the interferometric phase estimation.

\acknowledgments
LTK, IHLG, MAL acknowledge financial support from the Argentine funding agencies CONICET and ANPCyT.
GMB is partiallty supported by the Fondazione di Sardegna within the project ``Strategies and Technologies for Scientific Education and Dissemination''.

\appendix
\section{Quantum Fisher information}
\label{app:QFIproof}

For unitary maps $U_\phi = \exp(-i\phi H)$, where $H$ is the corresponding generating hermitian operator, the QFI does not depend on $\phi$ and it can be written as (see \eg \cite{Paris2009,Toth2014})
%
$\qfi(\rho^\inp,H) = 2 \sum_{i,j} \frac{\lambda_i - \lambda_j}{\lambda_i + \lambda_j} \braket{e_i|H|e_j}$,
%
where  $\{\ket{e_k}\}$ and $\{\lambda_k\}$ are the eigenstates and eigenvalues, respectively, of $\rho$.
In particular, when the initial state is pure $\rho^\inp = \ket{\Psi^\inp}\bra{\Psi^\inp}$, the QFI is proportional to the variance of the hamiltonian in the initial state, that is,
%
$  \qfi(\ket{\Psi^\inp},H) = 4 \Delta^2(H)$. 
%
In our case $H= J_\mu \otimes I + I \otimes J_\nu$, so that
\begin{equation}
\begin{aligned}
\label{eq:QFIpureU}
 & \qfi(\ket{\Psi^\inp},H) = 4\Delta^2(J_\mu \otimes I) + 4\Delta^2(I \otimes J_\nu) +\\
 & 8 \left(\braket{\Psi^\inp |J_\mu \otimes J_\nu|\Psi^\inp}-\braket{\Psi^\inp |J_\mu \otimes I|\Psi^\inp} \right) \braket{\Psi^\inp |I \otimes J_\nu|\Psi^\inp}.
  \end{aligned}
\end{equation}
For the initial state given in~\eqref{eq:input}, it can be shown that the last term, which are the covariance between $J_\mu \otimes I$ and $I \otimes J_\nu$ in the initial state, are zero.
On the other hand, the first two terms reduce to
\begin{widetext}
\begin{equation}
 \begin{aligned}
 & 4\Delta^2(J_\mu \otimes I)  = 4\braket{\Psi^\inp |J^2_\mu \otimes I|\Psi^\inp} = 2n (n+1) \I^n + \sum_{k=1}^{n-1} \frac{n!}{k!(n-k)} 2n (n-k+1) \I^{n-k} (1-\I)^k + n (1-\I)^n,  \\
& 4\Delta^2(I \otimes J_\nu) = 4\braket{\Psi^\inp |I \otimes J^2_\nu|\Psi^\inp}=\sum_{k=1}^{n-1} \frac{n!}{k!(n-k)} k \I^{n-k} (1-\I)^k + n (1-\I)^n.
\end{aligned}
\end{equation}
\end{widetext}

After some algebra and using that $\sum_{k=1}^{n-1} \frac{n!}{k!(n-k)} \I^{n-k}\I^k= 1-\I^n - (1-\I)^n$ and $\sum_{k=1}^{n-1} \frac{n!}{k!(n-k)} k \I^{n-k} (1-\I)^k= n (1-\I) (1-(1-\I)^{n-1})$ Eq.~\eqref{eq:QFI} is obtained.

Eq.~\eqref{eq:QFImixed} can be obtained in a straigthforward way by using the fact that the QFI of a mixed state of the form $\rho^\inp= (1-\epsilon) \ket{\Psi^\inp}\bra{\Psi^\inp} + \epsilon \frac{\mathbb{I}}{d}$ can be written as~\cite{Hyllus2012,Toth2012}
\begin{equation}
\qfi(\rho^\inp, H) = \frac{(1-\epsilon)^2}{1-\epsilon+\epsilon/d}  \qfi(\ket{\Psi^\inp}, H).
\end{equation}

\section{Noiseless and noisy theoretical probabilities}
\label{app:prob}

In the ideal case one has to be able to measure ten probabilities from the output state~\eqref{eq:out2phtons} to obtain the QFI.
The theoretical values of these noiseless probabilities are given by
%
\begin{align}
& p(2_h000|\phi)= p(02_v00|\phi) = \I(\varphi) \frac{\sin^2(\phi)}{2}, \\
& p(1_h1_v00|\phi) = \I(\varphi) \cos^2(\phi), \\
& p(1_h01_h0|\phi)= p(01_v01_v|\phi) = (1-\I(\varphi)) \frac{\sin^2(\phi)}{4}, \\
& p(1_h001_v|\phi)= (1-\I(\varphi)) \cos^4(\phi/2), \\
& p(01_v1_h0|\phi) = (1-\I(\varphi)) \sin^4(\phi/2), \\
& p(002_h0|\phi)= p(0002_v|\phi) = p(001_h1_v|\phi) =0.
\end{align}
%
Therefore, only seven out of the ten probabilities are needed.
From these probabilities, the Fisher information is computed as usual, $F(\phi)=  \sum_{m \in \mathfrak{M}} \frac{\left|\frac{dp(m|\phi)}{d\phi}\right|^2}{p(m|\phi)}= 2 \I(\varphi) + 2$ with $\mathfrak{M} = \{2_h000,02_v00,1_h1_v00,1_h01_h0,01_v01_v,1_h001_v,01_v01_v,\allowbreak 001_h1_v,002_h0,0002_v \}$.
By simple inspection it becomes clear that this is the Quantum Fisher Information, that is: $\qfi(\ket{\Psi^\inp}, H) = F(\phi)$.

Notice that if only double or coincident events are detected, that is, modes $a$ and $b$ are not resolved (or $\mu$ and $\nu$ in the general scheme) as in previous works~\cite{Birchall2016,Jachura2016}, the theoretical values of  measurement probabilities are
\begin{align}
\tilde{p}(2_h0|\phi) &= p(02_v|\phi)= (1+\I(\varphi)) \frac{\sin^2(\phi)}{4} \\
\tilde{p}(1_h1_v|\phi) &= \frac{1}{4} \left[3-\I(\varphi)+(1+\I(\varphi)) \cos(2\phi) \right].
\end{align}

%
%
%
This leads to a phase-dependent FI:
$\tilde{F}(\phi) =  \sum_{m \in \mathfrak{\tilde{M}}} \frac{\left|\frac{d\tilde{p}(m|\phi)}{d\phi}\right|^2}{\tilde{p}(m|\phi)}= \frac{4 [1 + \I(\varphi)] [1 + \cos(2\phi)]}{3 - \I(\varphi) + [1 + \I(\varphi)] \cos(2\phi)}$ where $\mathfrak{\tilde{M}} = \{2_h0,02_v,1_h1_v \}$.
Notice that $\tilde{F}(\phi) \leq \qfi(\ket{\Psi^\inp}, H)$ with equality at phases $\phi=2k \pi$.

In a more realistic case where the state preparation is affected by white noise, modeled by the initial state $\rho^\inp= (1-\epsilon) \ket{\Psi^\inp}\bra{\Psi^\inp} + \epsilon \frac{I}{10}$ with $\ket{\Psi^\inp}$ given by Eq.~\eqref{eq:initial2phtons}, the probabilities change as $p'(m|\phi)=(1-\epsilon)p(m|\phi)+\epsilon\frac{1}{10}$ for the same projective measurement $\M$ as performed for the initial pure state case.
It can be shown that the Fisher information from these noisy probabilities depends on the phase as it is shown in Eq.~\eqref{eq:FIruido}.
Moreover, the FI~\eqref{eq:FIruido} is strictly lower than the corresponding QFI (c.f. Eq.~\eqref{eq:QFImixed} with $n=1$ and $d=10$) for any amount of noise.
This means that in the noisy case, the projective measurement  $\M$ is no longer the optimal one.
If one considers only measurements of double or coincident events, the FI is even lower.
Therefore, it is interesting to study which is the optimal measurement in the noisy case, at least from the theoretical point of view.
Since this is a hard problem in general, we focus on the extreme case of highest degree of indistinguishability $\I=1$.
From the projective measurement $\M''=\{\ket{m''}\bra{m''}\}$ with
$\ket{m''} \in \{ \frac{\ket{2_h000} \pm \ket{02_v00}}{\sqrt{2}},\frac{\ket{1_h01_h0} \pm \ket{01_v01_v}}{\sqrt{2}},\frac{\ket{1_h001_v} \pm \ket{01_v1_h0}}{\sqrt{2}},\\ \ket{1_h1_v00}, \ket{001_h1_v},\ket{002_h0},\ket{0002_v} \}$, one obtains the FI $F''(\phi) = \frac{20 (1-\epsilon)^2 (4 \epsilon -5) \sin^2(2\phi)}{25 (1-\epsilon)^2 \cos^2(2\phi)- (4 - 5\epsilon)^2}$.
Interestingly enough, $F''(\pi/4+k\pi) = 4 \frac{(1-\epsilon)^2}{1-\frac{4}{5}\epsilon}$.
This means that $\M''$ is the optimal measurement for phases $\pi/4+k\pi$ for any amount of noise when $\I=1$. For other values of $\I$ the largest difference is of the order of $3\%$.
Nevertheless, the practical implementation of this measurement can be challenging.


\end{document}